\documentclass[twoside]{ilcws07}
\usepackage[latin1]{inputenc}
\usepackage[dvips]{graphicx,epsfig,color}
\usepackage{wrapfig,rotating}
\usepackage{amssymb,amsmath,array}

\begin{document}

% \title{H and A Discrimination with Linear Photon Polarization} %% 
\title{H and A Discrimination using Linear Polarization of Photons at the PLC}

\author{A.F.\.Zarnecki$^1$, P.Nie\.zurawski$^1$, M.Krawczyk$^2$\thanks{
%This work was partially supported by the Polish Ministry of Science
% and Higher Education, grant  no.~1~P03B~040~26.
MK acknowledges partial support by EU Marie Curie Research Training Network
HEPTOOLS, under contract MRTN-CT-2006-035505.}
%
% DO NOT MODIFY THE FOLLOWING '\vspace' ARGUMENT
\vspace{.3cm}\\
% Addresses and institutions (remove "1- " in case of a single institution)
1- Institute of Experimental Physics, University of Warsaw\\
   Ho\.za 69, 00-681 Warszawa, Poland 
%% Remove the next three lines in case of a single institution
\vspace{.1cm}\\
2- Institute of Theoretical Physics, University of Warsaw\\
   Ho\.za 69, 00-681 Warszawa, Poland\\
}

\maketitle

\begin{abstract}
First realistic estimate of the usefulness of the Photon Linear Collider 
with  linearly polarized photons as analyzer of the CP-parity of 
Higgs bosons is presented.
MSSM Higgs bosons H and A  with 300~GeV mass, 
for the model parameters corresponding 
to the so called ``LHC wedge'' region,
are considered.
When switching from circular to linear photon polarization
a significant increase in heavy quark production background, 
which is no longer suppressed by helicity conservation, and decrease
of the Higgs boson production cross sections by a factor of two is
expected. 
Nevertheless, after three years of Photon Linear Collider running 
heavy scalar and pseudoscalar Higgs bosons in MSSM can be distinguished 
at a 4.5~$\sigma$ level.
\end{abstract}

%**************************************************************

\section{Introduction}

The physics potential of a Photon Linear Collider (PLC) is very
rich and complementary to the physics program of the $e^+e^-$
and hadron-hadron colliders.
It is an ideal place to study the mechanism of the electroweak 
symmetry breaking (EWSB) and the properties of the Higgs sector, 
as it allows for a resonant production of the Higgs particles. 
In our previous studies we have performed realistic simulations of the Higgs 
boson production at PLC within the Standard Model~\cite{nzk_smbb,nzk_wwzz},
Two Higgs Doublet Model (2HDM)~\cite{nzk_2hdm}, MSSM~\cite{nzk_mssm} and
a generic model with the CP violating Higgs boson couplings~\cite{nzk_cp}.
In all cases we have assumed circular polarization of colliding photon beams, 
which is favourable from the point of view of production cross section 
and background suppression.
However, it does not allow to disentangle between production of heavy 
MSSM Higgs bosons H and A~\cite{nzk_mssm}.
It was suggested that for studies of the CP-parity of Higgs particles 
and search for a potential violation of the CP-invariance in the Higgs sector
linear photon polarization should be used~\cite{hagiwara}.
In the presented study~\cite{url} we have made the first realistic 
estimate of the Higgs measurement at PLC running with linearly polarized beams.

\section{Luminosity spectra}

\begin{figure}[t] 
\centerline{\includegraphics[width=5.5cm]{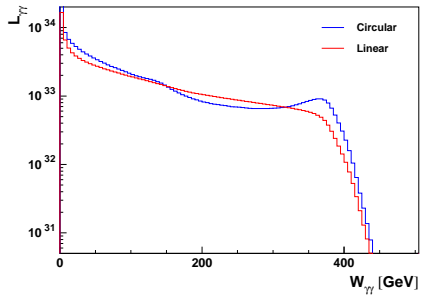}}
 \caption{ Expected luminosity spectra at the PLC 
            for circular and linear photon beam polarizations,
           as obtained from CAIN simulation. Parameters of the 
           TESLA Photon Collider were used with electron 
           beam energy of 250 GeV and electron circular polarization of 85\%.}  
  \label{fig:linlumi}  
\end{figure}

\begin{figure}[t]  
\centerline{
\includegraphics[width=5.5cm]{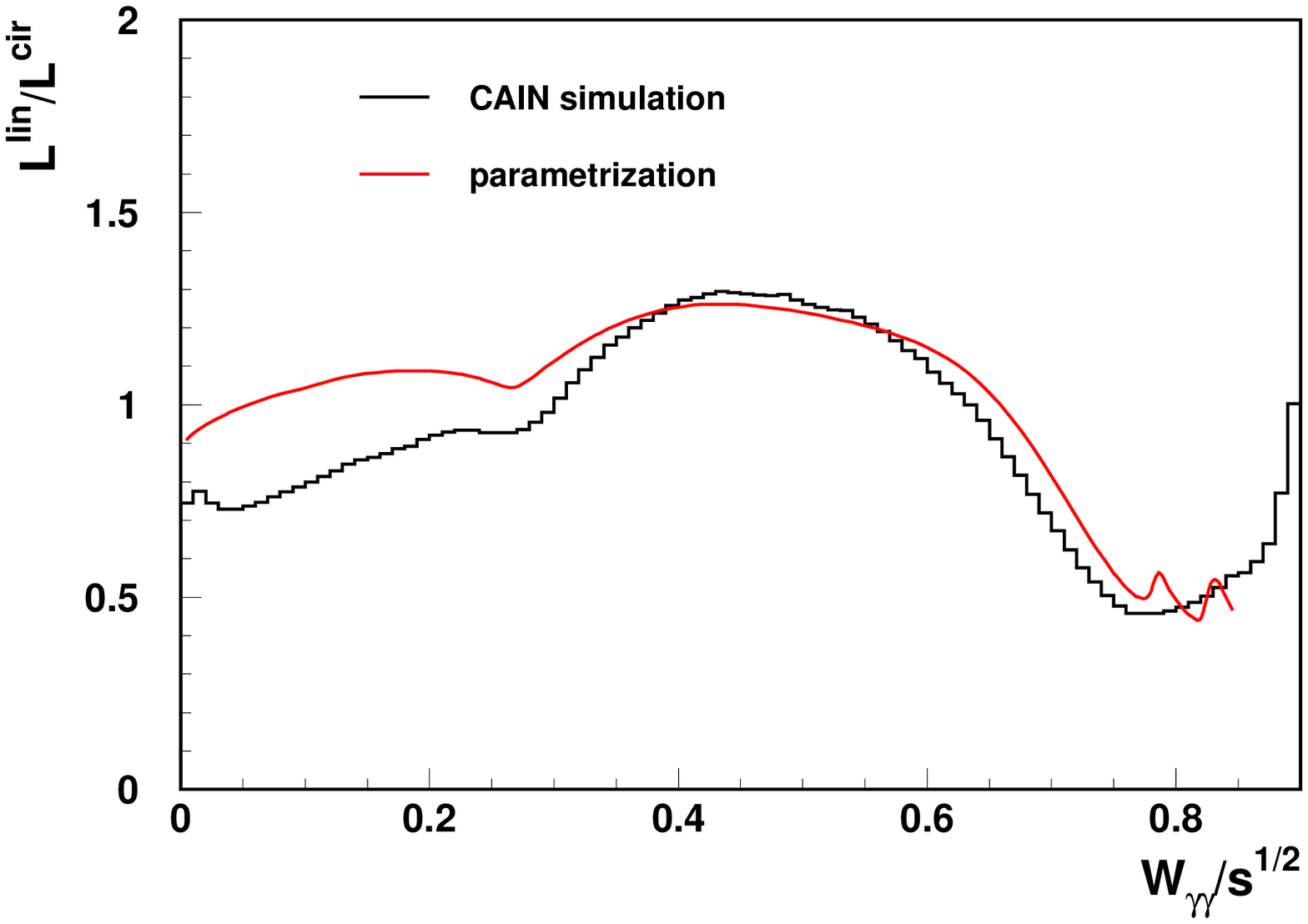}
\hspace{1cm}
\includegraphics[width=5.5cm]{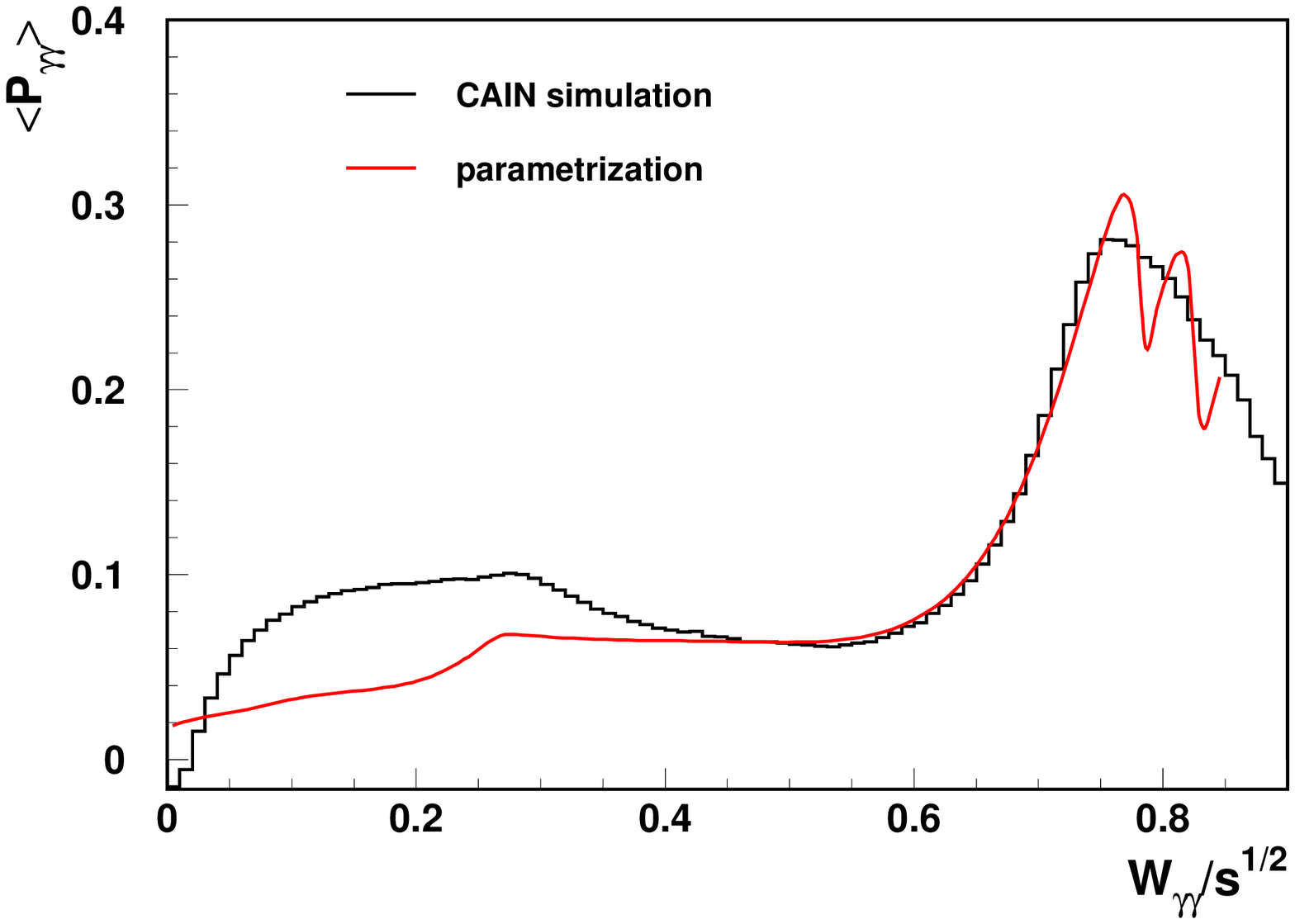}
}
  \caption{ Comparison of the CAIN simulation results for linear 
           photon beam polarization with new luminosity spectra
           parametrization based on CompAZ. Ratio of $\gamma \gamma$
           luminosities for linear and circular polarization (left plot)
           and the average product of polarizations for colliding photons
           (right plot) are considered.}  
  \label{fig:linfit}  
 \end{figure}

In order to perform this analysis, parametrization of the PLC 
luminosity spectra CompAZ~\cite{compaz} has to be extended to
linear photon polarization.
As pointed out in~\cite{vtlin}, there are significant correlations between
polarizations of colliding photons, resulting in sizable increase of the
effective polarization. 
They are of special importance at high beam energies, where average 
beam polarizations are low. Only due to this phenomena measurements
with linear polarization are possible at all in this energy region.
CAIN~\cite{cain} program was used to simulate Compton backscattering process
with linearly polarized laser photons at the PLC, taking into
account all correlations.
Based on this simulation we derive expected luminosity 
and polarization in $\gamma \gamma$ collisions.
Figure \ref{fig:linlumi} shows the comparison of the expected PLC luminosity 
spectra for circular and linear beam polarizations.
For linear polarization luminosity spectra is no longer peaked at high 
energies and the luminosity in the region of high energy $W_{\gamma \gamma}$ 
decreases.
Simulation results were used to constrain parameters describing 
polarization gain in $\gamma \gamma$ collisions.
As shown in Figure \ref{fig:linfit} the obtained parametrization 
properly describes modification of the luminosity 
spectra due to change of beam polarization,
especially in the high energy domain.
Also the average product of photon polarizations,
$\langle P_{\gamma\gamma} \rangle$, is well described.
In the region of large $W_{\gamma \gamma}$, 
 $\langle P_{\gamma \gamma} \rangle$
of up to about 30\% can be obtained.

\section{Cross section measurement}

\begin{figure}[t] 
\centerline{
\includegraphics[width=5.5cm]{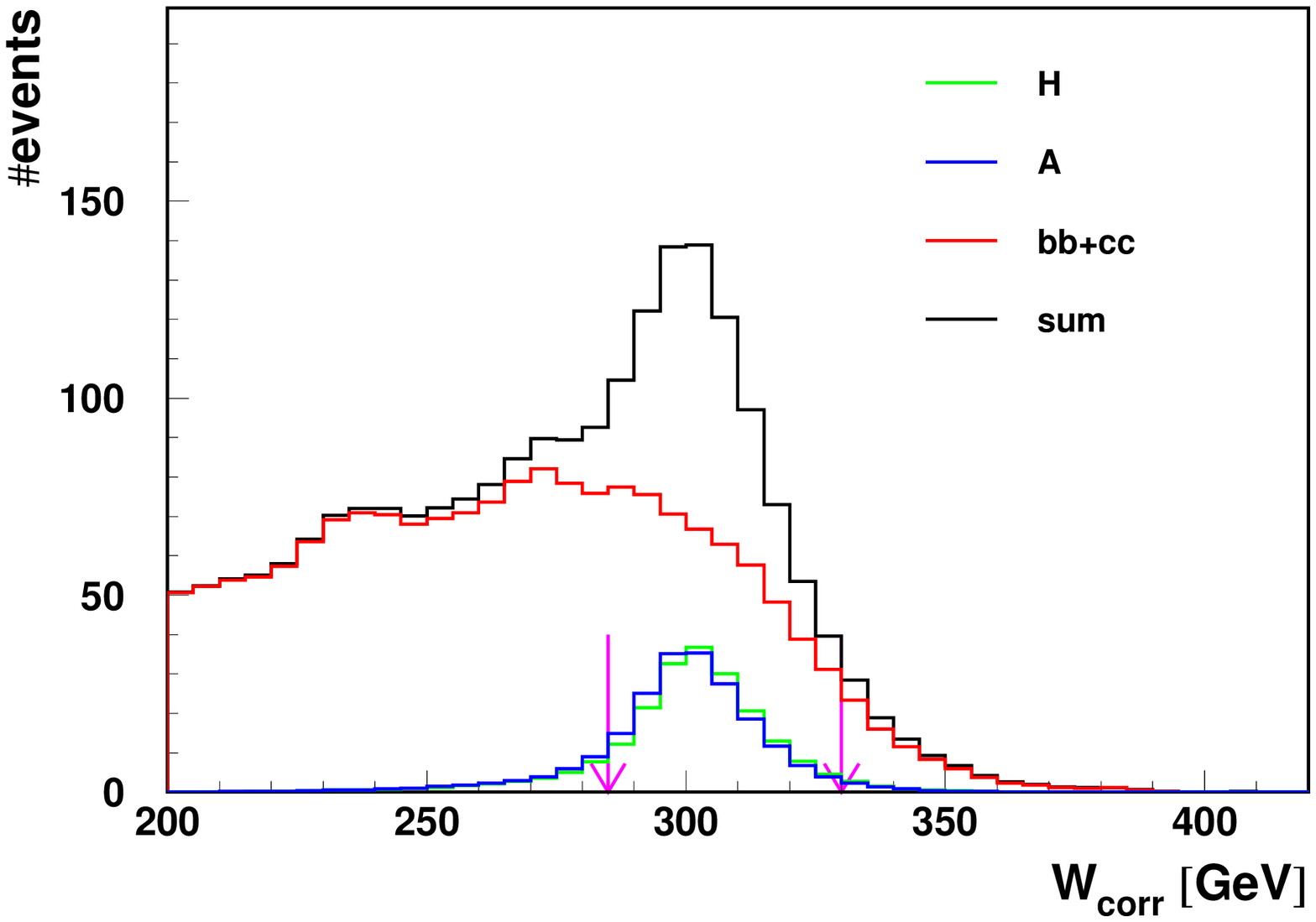}
\hspace{1cm}
\includegraphics[width=5.5cm]{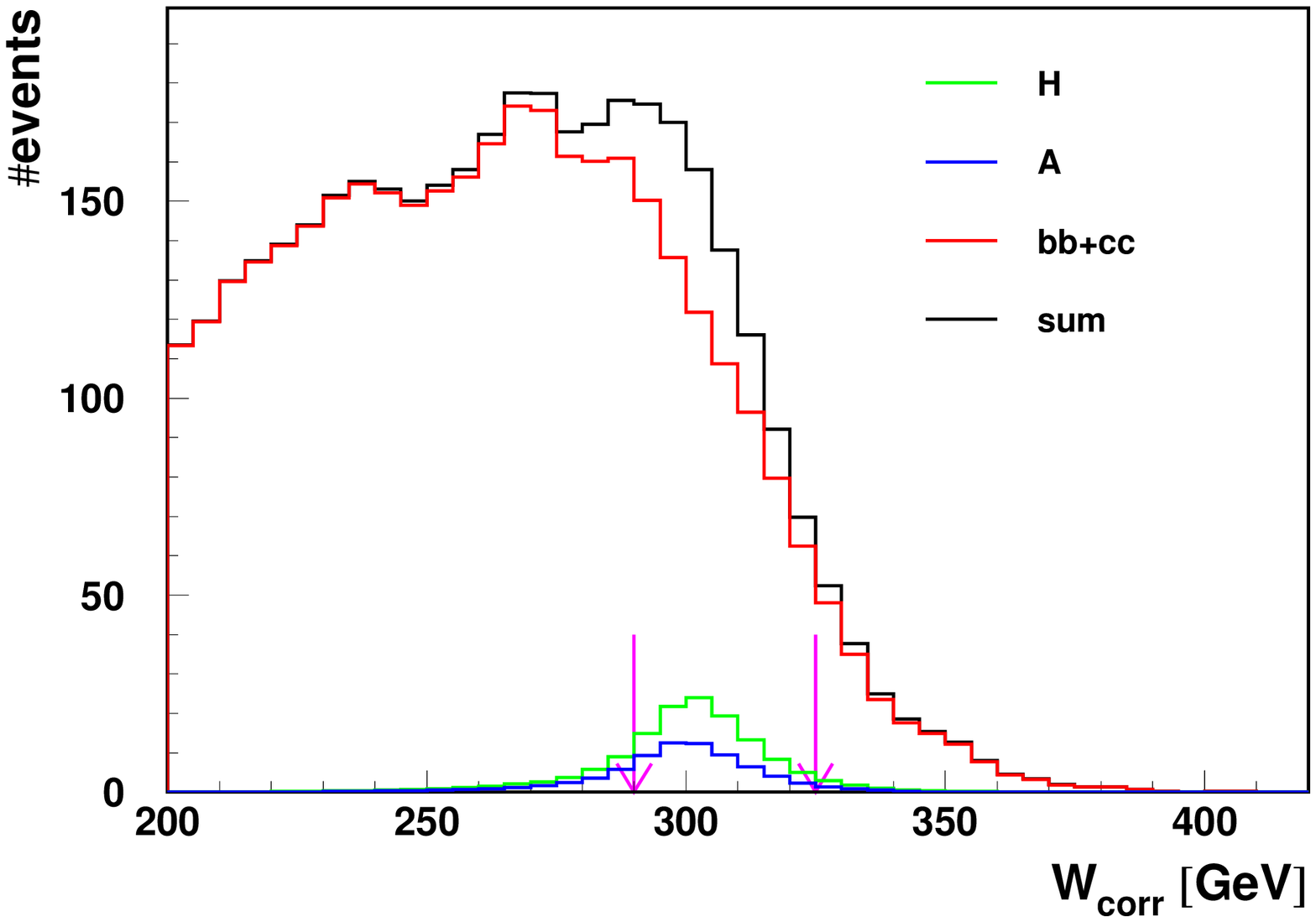}
}
  \caption{ Reconstructed invariant mass distributions
expected after one year of PLC running, for circular (left) 
and linear (right) photon beam polarizations. }  
  \label{fig:wcorr}  
 \end{figure}

Parametrized luminosity spectra are used to simulate Higgs boson
production with linearly polarized beams.
We considered production of the MSSM Higgs bosons H and A for
the parameter values corresponding to the  ``LHC wedge'' region:
$M_A =$ 300 GeV, $\tan \beta = 7$, $M_{2} = \mu = $ 200 GeV. 
For these parameter values, bosons H and A are almost degenerate
in mass and can not be distinguished on the detector level.
Analysis follows our previous study described in~\cite{nzk_mssm}.
However, here only the background from heavy quark production 
$( \gamma \gamma \rightarrow Q \bar{Q} (g)$, $Q=b,c$)
is taken into account.
Figure~\ref{fig:wcorr} shows the reconstructed invariant mass distributions
expected after one year of PLC running, for circular and linear
beam polarizations.
We observe that with linear polarization signal (H and A production) decreases
by about factor of two. 
This is because of the luminosity drop at high $W_{\gamma \gamma}$, but
also due to the reduced circular polarization of the photon beam
(even with 100\% linear laser polarization, some degree of the circular  
polarization is expected due to the polarization of the incident electron beam).
Smaller degree of circular polarization results also in significant 
increase in heavy quark production background, which is no longer suppressed
by the helicity conservation.
After independent cut optimization, signal to background ratio for 
linear polarization is about factor of 3 smaller than for the 
circular polarization.
Precision of the cross section measurement, for H and A production,
changes from about 8\% for circular beam polarization to about 18\% for 
linear one, after one year of PLC running.

\section{Discrimination between H and A}

\begin{figure}[t] 
\centerline{
\includegraphics[width=5.5cm]{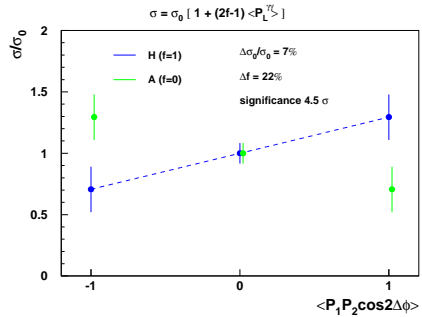}
}
  \caption{ Expected precision of Higgs boson production cross section
 measurements after three years of PLC running with circular 
and two linear laser beam polarizations. }  
  \label{fig:fits}  
 \end{figure}

For linear photon polarization we observe a clear difference between 
production of scalar and pseudo-scalar Higgs bosons, as shown in 
Figure~\ref{fig:wcorr} (right). 
By selecting parallel or perpendicular orientation of linear polarizations
of two beams, we enhance production of scalar or pseudo-scalar state,
respectively.
By combining measurements with different polarization orientations,
cross sections for $H$ and $A$ production can be disentangled.
After three years of PLC running (one year with each orientation of
linear polarization and one year with circular polarisation) cross
section determination precision of 22\% can be obtained.
Hypothesis of pure scalar or pure pseudo-scalar production (assuming
that the total production cross section for unpolarized beam is the same
as in the considered MSSM scenario)
can be distinguished at 4.5~$\sigma$ level, see Figure~\ref{fig:fits}.
Polarization of the photon beam obtained in the process of Compton 
backscattering is determined by polarizations of the incident laser
and electron beams.
Circular and linear polarization configurations considered so far
correspond to the laser light with 100\% circular
or 100\% linear polarization, and  85\% circular 
polarization for electrons.
However, one could also consider laser with mixed polarization.
In fact, highest  $\langle P_{\gamma \gamma} \rangle$
can be obtained by using 95\% linear laser polarization with additional 
contribution of 30\% circular laser polarization. 
% opposite to the polarization of the electron beam.
%
However, the corresponding polarization configuration results 
also in sizable decrease of luminosity at high $W_{\gamma \gamma}$ 
so that the final measurement precision is significantly worse.
Therefore we can conclude that the best separation between $H$ and $A$ states
can be obtained with 100\% linear laser polarization.

\section{Conclusions}

We presented the first realistic estimate of the Higgs boson 
CP-parity determination at the PLC
with  linear beam polarization.
Heavy MSSM Higgs bosons H and A, for model parameters corresponding 
to the so called ``LHC wedge'', were considered.
Significant increase in heavy quark production background, 
which is no longer suppressed by helicity conservation, and decrease
of the Higgs boson production rate 
result in the cross section measurement precision much lower than
for the circular beam polarization.
Nevertheless, after three years of PLC running 
cross sections  for $H$ and $A$ production can be separately
measured with precision of about 20\%.
Hypotheses of pure scalar or pure pseudo-scalar nature of the Higgs boson
(assuming the same value for the total production cross section)
can be distinguished at 4.5~$\sigma$ level.

% ****************************************************************************
% BIBLIOGRAPHY AREA
% ****************************************************************************

\begin{footnotesize}

\end{footnotesize}

% ****************************************************************************
% END OF BIBLIOGRAPHY AREA
% ****************************************************************************

\end{document}